\documentclass{PoS}

\usepackage{graphicx}
\usepackage{float}
\usepackage{subfig}
\usepackage{multirow}

\title{Coherent vector meson production at an electron ion collider}

\ShortTitle{Coherent vector meson production at an electron ion collider}

\author{\speaker{M. Lomnitz}\\
       Lawrence Berkeley National Laboratory\\
        E-mail: \email{mrlomnitz@lbl.gov}}   

\abstract{Exclusive vector meson electroproduction over a broad $Q^2$ range offers a unique opportunity to probe the gluon structure of nuclei to measure nuclear shadowing, and to search for gluon saturation and/or the colored glass condensate at an Electron-Ion Collider.   Understanding the kinematic distributions and cross sections for specific processes can impact detector design to maximize their acceptance and strengthen the physics case. We will discuss predictions from a Monte Carlo generator eSTARlight, a tool developed to study production of exclusive vector meson final states. We present final state distributions and production rates for the $\rho$, $\phi$, J/$\psi$, $\psi'$ and the $\Upsilon$ states in $ep$ and $eA$ collisions at the different energies. }

\FullConference{XXVI International Workshop on Deep-Inelastic Scattering and Related Subjects (DIS2018)\\
		16-20 April 2018\\
		Kobe, Japan}

\begin{document}
\bibliographystyle{JHEP}

\section{Introduction}
Coherent vector meson (VM) photoproduction ($Q^2 \approx 0$) and electroproduction (here $Q^2 > 1 \rm{GeV}^2$) are important channels \cite{Accardi:2012qut} for investigating the parton structure of protons as well as nuclear targets.   In coherent VM production an incident photon fluctuates into a quark-anti quark ($q\bar{q})$ dipole which scatters elastically off a target and emerges as a real VM. The interaction occurs through the exchange of either a meson (at low photon energies) or a Pomeron (at higher photon energies).  The Pomeron is mostly gluonic, so exclusive VM production cross-sections can be related to the gluon structure functions \cite{Jones:2016ldq} using perturbative quantum chromodynamics. \\
These reactions have large cross-sections and simple final states, so they are expected to play an important role in the upcoming Electron-Ion Collider (EIC) which will systematically study production as a function of $Q^2$ and Bjorken-$x$.  Understanding the rates and kinematic distributions for photoproduction and electroproduction of VMs is essential to facilitate detector design and study the feasibility of the physics program at potential EICs.

\section{Modeling electroproduction in eSTARlight}
  In order to study the rates and kinematics of VM production at an EIC, it is necessary to model photon emission over a wide range in energies, track the outgoing electron and correctly model the cross-section and angular distributions over a wide range in virtuality $Q^2$. The generator, dubbed eSTARlight \cite{Lomnitz:2017}, uses the fact that the cross-section for production of a VM $V$ on a target $X$ can be written as follows:
\begin{equation}
\sigma(eX\rightarrow eXV) = \int dW \int \frac{dk}{k} \int dQ^2 \frac{d^2 N}{dk dQ^2} \sigma_{\gamma * X \rightarrow VX}(W,Q^2)
\end{equation}
where $k$ and $Q^2$ are the photon energy and virtuality respectively, $W$ is the $\gamma p$ center of mass energy, $d^2 N/dkdQ^2$ is the photon flux and $\sigma_{\gamma * X \rightarrow V X}$ is the $\gamma X$ cross-section.  In the frame of reference where the target is at rest (target frame), the photon flux in the equivalent photon approach (EPA) \cite{Budnev:1975} and is given by:
\begin{equation}
\frac{d^2 N}{dk dQ^2} \frac{dk}{k} dQ^2 = \frac{\alpha}{\pi}\frac{dk}{k} \frac{dQ^2}{Q^2}\left[ 1 - \frac{k}{E_e} + \frac{k^2}{2E_e^2} - \left( 1 - \frac{k}{E_e}\right)\left| \frac{Q_{min}^2}{Q^2}\right|\right]  
\end{equation} 
The photon flux is largest at small values of $k$ and $Q^2$.  The minimum $Q^2$ depends on the final state while the maximum is determined by the electron energy loss: $m_e^2k^2/E_e(E_e-k) < Q^2 < 4E_e(E_e-k)$.  To avoid the problematic endpoint region, we take $k <  E_e-10m_e$ with negligible loss. Furthermore, for the photon to interact coherently with the target, it must fluctuate into a $q\bar{q}$ dipole with lifetime $\tau > \hbar/2p_T$. To model this, we impose a requirement on the coherence length (time) of the boosted dipole $l_c = 2\hbar k/(Q^2+M_v^2)$ and ensure coherent production by imposing a minimum value for $k$ such that $l_c$ is greater than the nuclear radius. These effects are largest near threshold.  \\
  The cross-section for the $\gamma X \rightarrow VM X$ process uses parameterizations to existing HERA data \cite{Adlo:2000,Aaron:2010}. For proton targets we assume the the following functional form:
\begin{equation}
\sigma_{\gamma p \rightarrow V p } = \left(\frac{M^2_V}{M_V^2 +  Q^2}\right)^n \sigma(W,Q^2 = 0 ) f(M_V)
\end{equation}
where the power $n=c_1 +c_2(Q^2+M_V^2)$ are taken from fits \cite{Klein:2016yzr}  and $f(M_V)$ is typically a Breit-Wigner, but may include additional components such as continuum direct $\pi^+\pi^-$ production. $\sigma(W,Q^2=0)$ is the cross-section for VM photo-production with real photons and can be parametrized as the sum of two power laws:
\begin{equation}
\sigma(W,Q^2=0)  = \sigma_P W^\epsilon + \sigma_M W^\eta
\end{equation}
where the first term describes production via Pomeron exchange and the second via meson exchange, and the values of the 4 parameters $\sigma_P,\epsilon,\sigma_M$ and $\eta$ depend on the final state particle \cite{Klein:1999}. For light particles the cross section rises slowly with $W$ while for heavier VM species the value of $\epsilon$ is larger and a more complicated expression is used to account for near threshold effects \cite{Aaron:2010}.   Production via meson exchange occurs only for the lightest species ($\rho$ and $\omega$), contributing mostly to production at low photon energies. \\
For heavier targets we utilize a quantum Glauber calculation \cite{Frankfurt:2002} which accounts for the possibility of a single incident $q\bar{q}$ interacting multiple times with the target, especially important for large dipoles, i.e. light vector mesons. The procedure follows \cite{Klein:1999}, but replaces it's Eq. (12) with the following:
\begin{equation}
\sigma_{tot}(VA) = 2\int d^2\vec{r} \left[1 - e^{0.5\sigma_{tot}(Vp) T_{AA}(\vec{r})}\right]
\end{equation}
where $\sigma_{tot}(VA)$ is the total vector meson-nucleus cross-section, $\sigma_{tot}(Vp)$ is the total vector meson-proton cross-section, $T_{AA}(\vec{r})$ is the nuclear thickness function calculated for a Woods-Saxon density distribution. 
It is also important to accurately model the VM decay products angular distributions to study detector acceptance. In the limit $Q^2\rightarrow 0$ the photons are all polarized transverse to the beam direction. The VM retains the photon spin state, the angular distributions of the daughter particles is determined from the Clebsh-Gordon coefficients. However, virtual photons may also be longitudinally polarized and the description becomes considerably more complicated and the evolution with $Q^2$ is rather complex. Because of this, we describe the VM decay in the helicity system and parametrize the $Q^2$ dependence of the longitudinal-to-transverse cross-section ratio ($R_V$) following the procedure in \cite{Schildknecht:1999} and extract the VM spin matrix ($r$) elements which can then be used to establish the angular distributions.\\
  Rejection sampling is used to generate the relevant kinematic variables by sampling from all of the distributions previously mentioned. Look-up tables are created when initializing the generator to speed up event generation. 
\section{Results: cross-sections and kinematic distributions}
\begin{figure*}
\center
\subfloat{
	\includegraphics[width=0.45\textwidth]{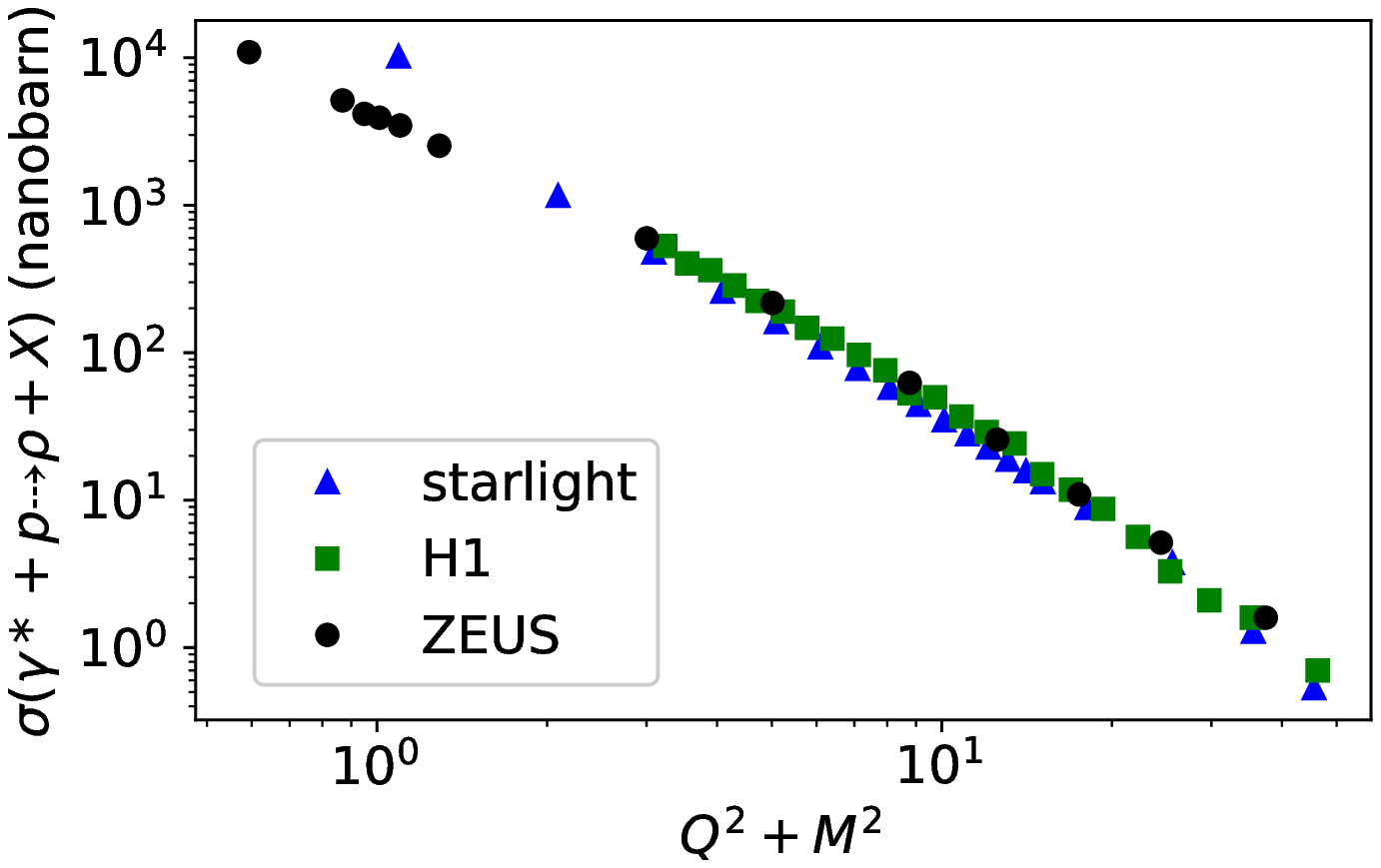}}
\subfloat{
	\includegraphics[width=0.45\textwidth]{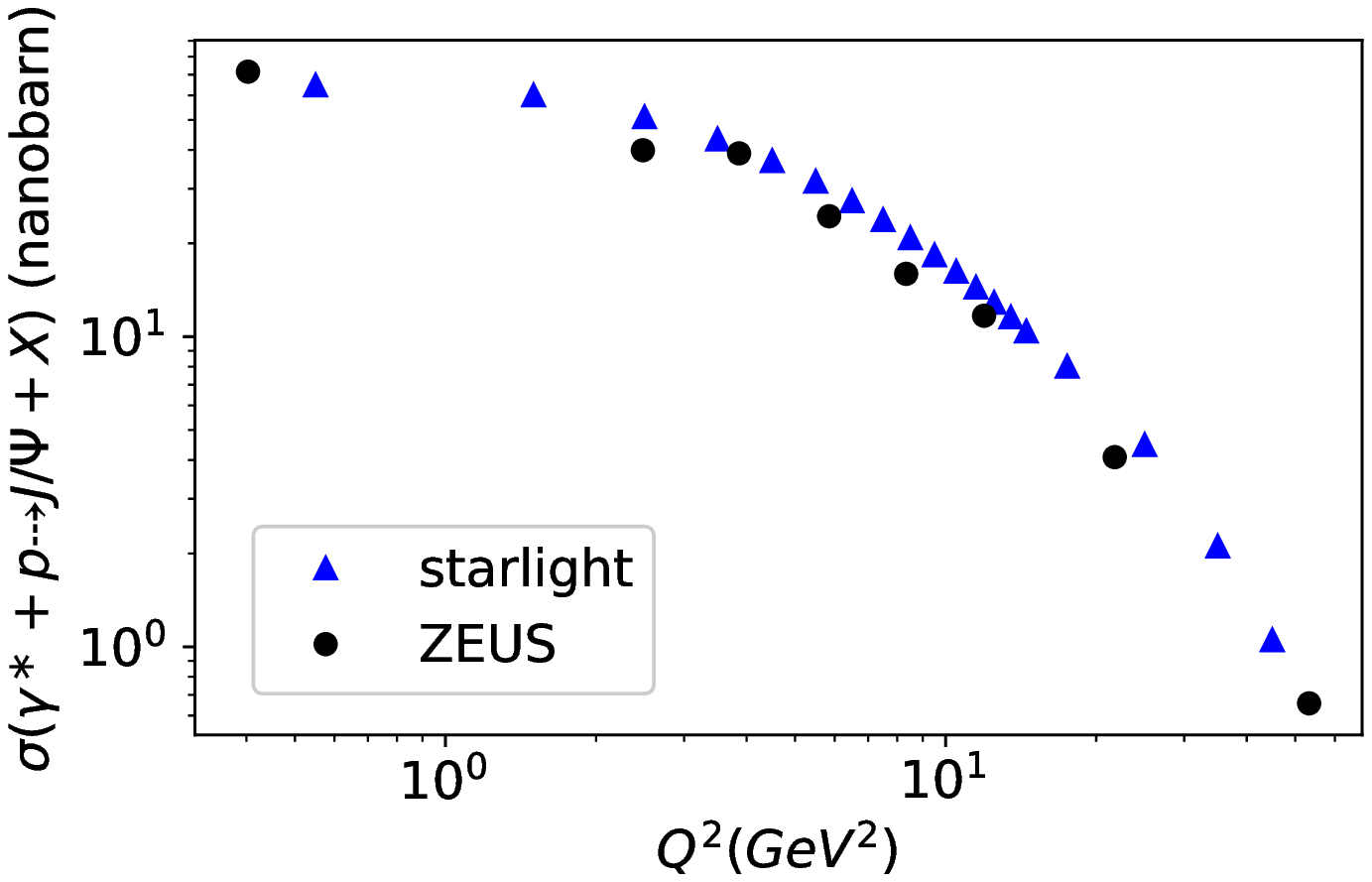}}
\caption{Calculations of photo-nuclear cross sections $\sigma(\gamma p \rightarrow Vp)$ vs. $Q^2$ compared with HERA data.  The $x$ axes are chosen to match that from the experimental data shown.}
\label{fig:hera_vs_q2}
\end{figure*}

\begin{figure*}
\center
\subfloat{
	\includegraphics[width=0.45\textwidth]{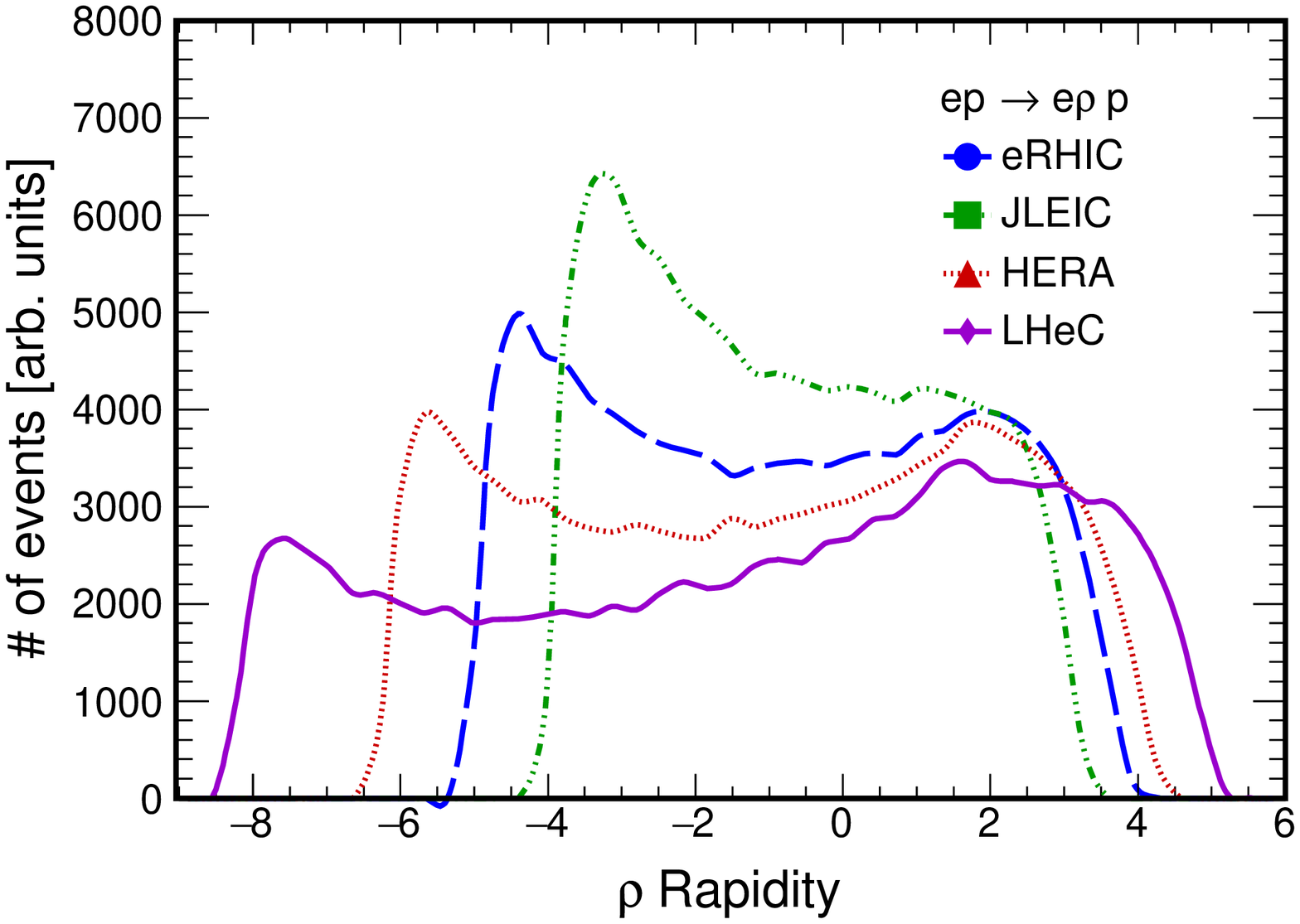}}
\subfloat{
	\includegraphics[width=0.45\textwidth]{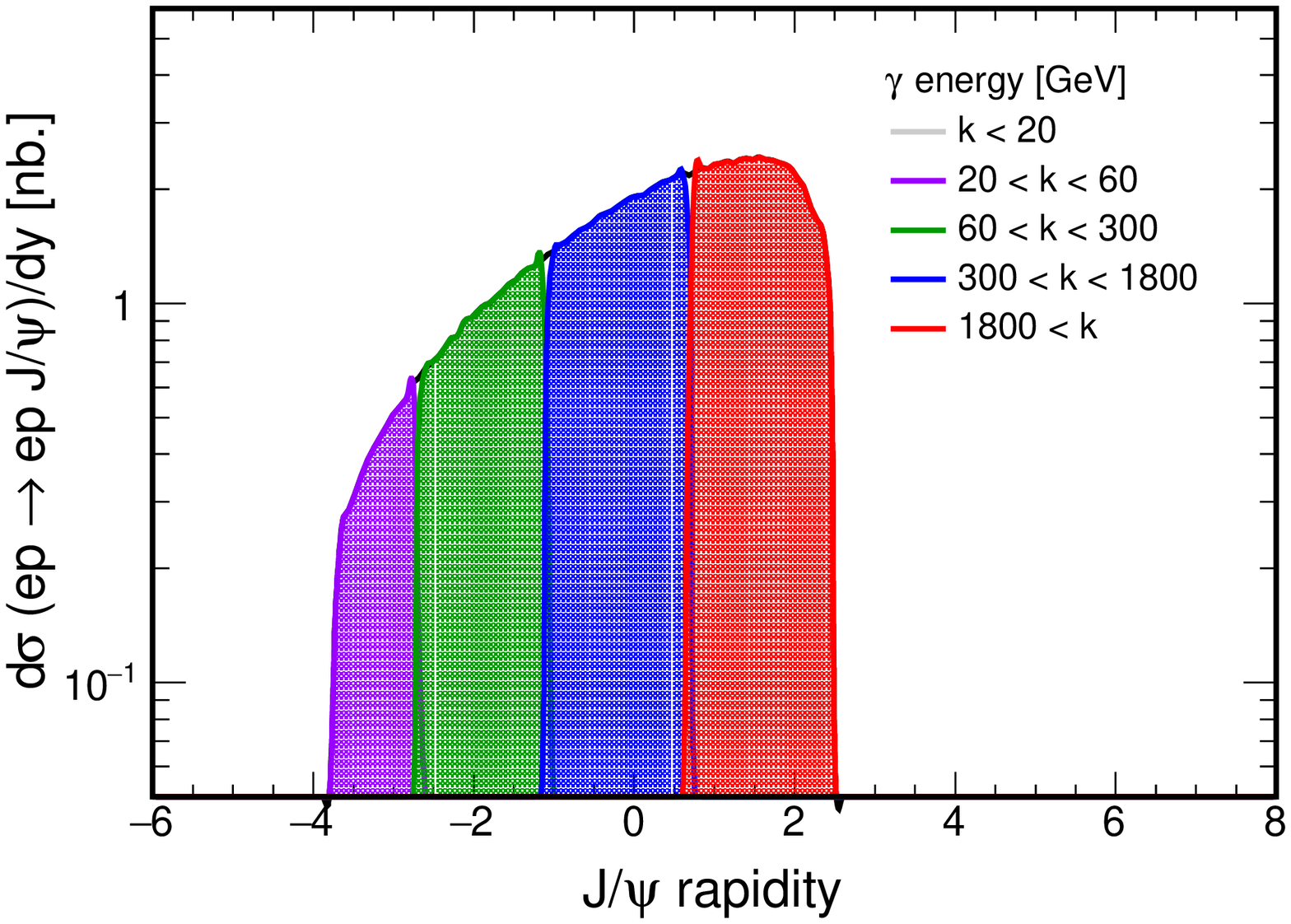}}
\caption{Predictions for the rapidity distribution of vector mesons produced in $ep$ collisions. Left: dN/dy for $\rho$ at the different proposed facilities and HERA for comparison. Right: J/$\psi$ production at eRHIC. The different band colors show the photon energy in the target frame.  }
\label{fig:ep_rapidities}
\end{figure*}

\begin{figure*}
\center
\subfloat{
	\includegraphics[width=0.45\textwidth]{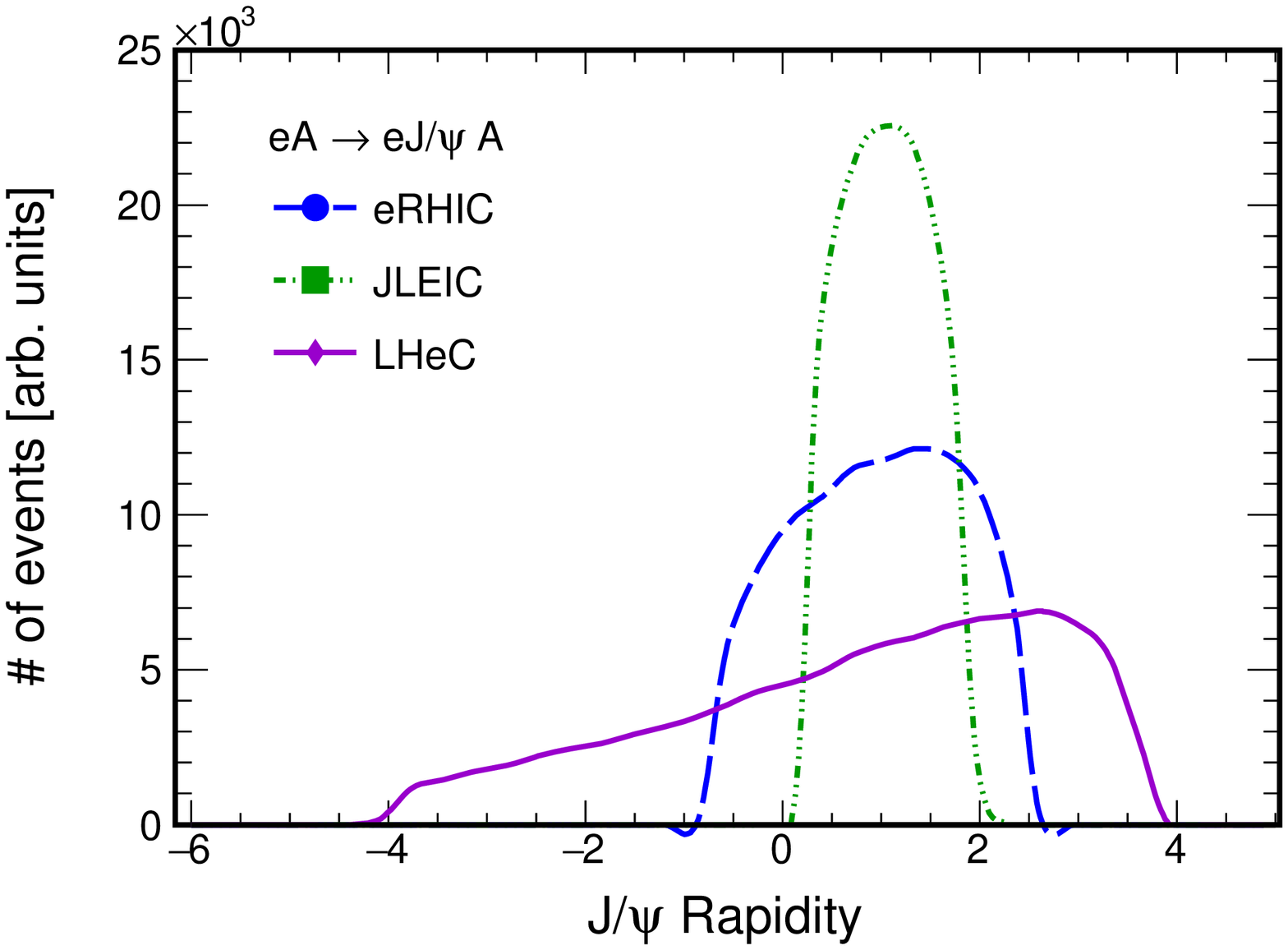}}
\subfloat{
	\includegraphics[width=0.45\textwidth]{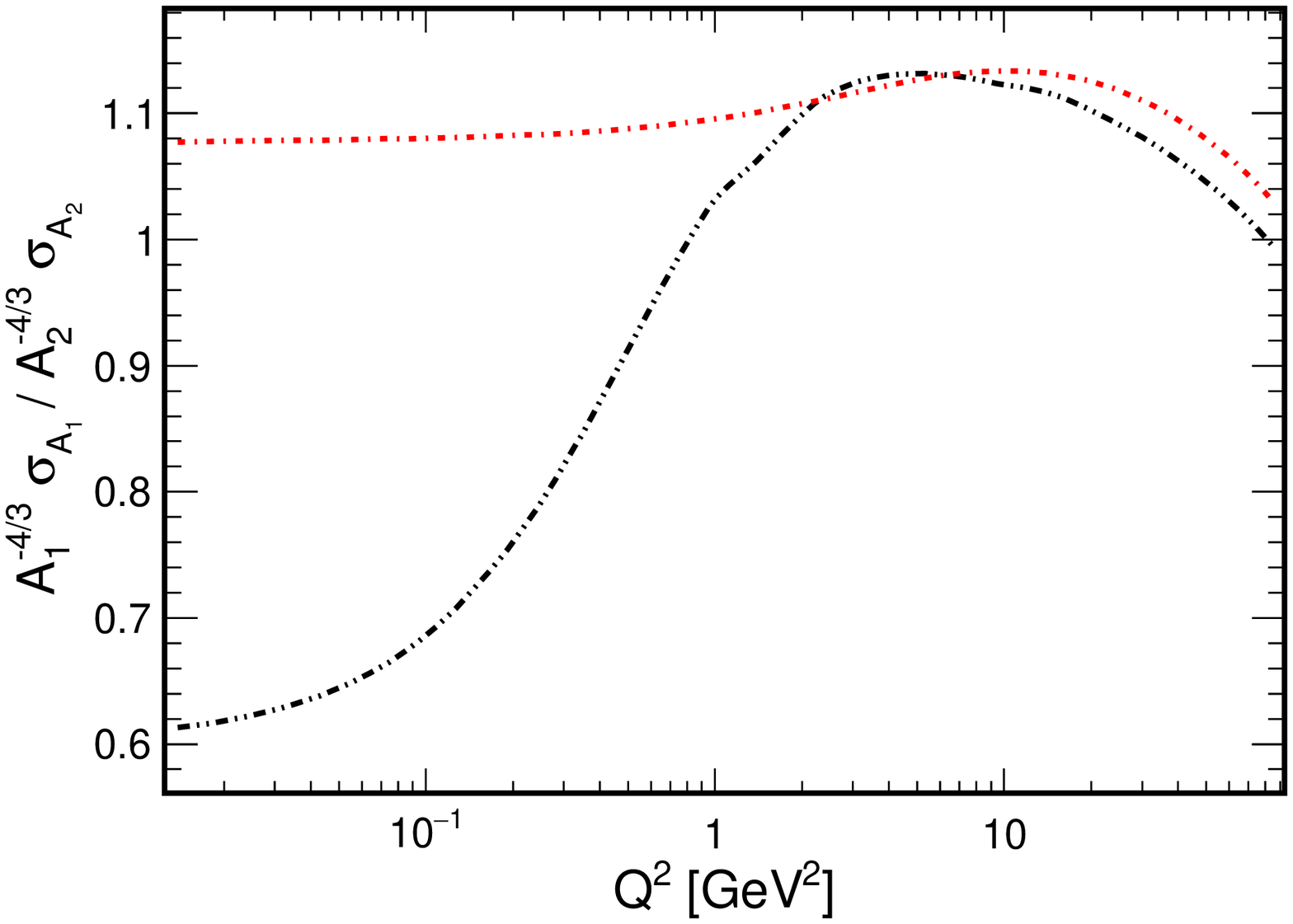}}
\caption{Predictions for photoproduction of vector mesons in $eA$ collisions. Left: dN/dy for $\rho$ at the different proposed facilities. Right: the scaled cross-section ratio on gold vs. iron targets for both $\rho$ and J/$\psi$. \label{fig:eA_collisions}}
\end{figure*}

We will present calculations for the photo-production ($Q^2 < 1$ GeV$^2$) and electro-production ($Q^2 > 1$ GeV$^2$) for a variety of VM species in both $ep$ and $eA$ collisions at a variety of proposed Electron-Ion Colliders, as well as at HERA to cross check our results.  Table \ref{tab:accel} summarizes the characteristics for the accelerators included in these studies. 
\begin{table}
\centering
\begin{tabular}{ | l | c | c | c |}
\hline
 Accelerator & Electron Energy & Proton Energy & Heavy Ion Energy \\
\hline
eRHIC         & 18 GeV  & 275 GeV & 100 GeV/A\\
\hline           
JLEIC         & 10 GeV  & 100 GeV  & 40 GeV/A\\ 
\hline
LHeC         &  60 GeV  &   7 TeV & 2.8 TeV/A \\ 
\hline
HERA         &                    27.5 GeV          &   920 GeV  & -  \\  
\hline
\end{tabular}
\caption{The characteristics for the accelerators considered here. The ion energies are per-nucleon. JLEIC and LHeC studies assume lead beams, while eRHIC uses gold. 
\label{tab:accel}
}
\end{table}
Figure \ref{fig:hera_vs_q2} compares eSTARlight results with existing data from the H1 and ZEUS experiments, showing the photo-nuclear cross sections $\sigma( \gamma p \rightarrow V p )$ vs $Q^2$ for (left) $\rho$ and (right) $J/\psi$. The data \cite{Aaron:2010,Chekanov:2004} and calculations agree over a wide range in photon virtuality.\\
  Tables (III) and (IV) in \cite{Lomnitz:2017} show the expected rates of photo- and electro-production of a series of VM species at the different EIC's for both proton and heavy ion targets.  Although the cross-sections for heavy ions are considerably larger, they rates are offset by the decrease in the per-ion luminosity. The quantum Glauber approach for heavy ions neglects more complex nuclear effects, such as shadowing, and may result in overestimates for the rates in heavy ions. The rates for electro-production are roughly 100 times lower that for photo-production, a consequence of the lower photon fluxes at high virtuality together with the the decrease in the photon-nucleon cross-section with increasing $Q^2$. This decrease is larger for heavy ions due to the lower per nucleon energy in the collision which further reduces the production of high $Q^2$ photons.\\
  Figure \ref{fig:ep_rapidities} shows (left) the rapidity distributions of produced $\rho$ at the 3 EIC's studied, as well as at HERA.  The individual curves have not been scaled by the cross-sections at the different colliders in order to facilitate the comparison on a single plot.  Production occurs over a wide rapidity window, roughly matching to the energy of the exchanged photon. The peak at large negative rapidities corresponds to production via meson exchange (not present in heavier VM species) while the peak at large positive rapidities is from photons with large energies. This is further illustrated in the panel on the right, which shows the rapidity distribution of produced J/$\psi$ at eRHIC. The different color bands indicate the energy of the photon, in the target frame.  These distributions can already inform on design aspects of potential EIC's.  Given that an overlap with data collected from CEBAF is desirable, this would require instrumentation at large negative rapidities or, alternatively, running an EIC at lower center of mass energies.
Figure \ref{fig:eA_collisions} shows our predictions for production in $eA$ collisions. The panel on the left shows the rapidity distributions for J/$\psi$ production at the different colliders. Due to the lower center of mass energy per nucleon, the kinematic range available for VM production is reduced and the production occurs over a much narrower rapidity window. The right panel shows the evolution of the $\rho$ and J/$\psi$ cross-sections with $Q^2$. Following Ref. \cite{Mantysaari:2017}, we plot the ratio of the cross-section on gold and iron targets, scaled by $A^{-4/3}$.  In the absence of nuclear effects, this scaled ratio should be identical to 1. The ratio drops for low values of $Q^2$, demonstrating of shadowing effects in the nucleus. This effect is visible in the case of J/$\psi$ but is substantially smaller. Charm quarks are considerably heavier so the $c\bar{c}$ dipole is always small. For both species, the ratio for large values of $Q^2 > 5 $ GeV$^2$ is greater than one. In this regime the effect of multiple interactions by a single dipole is not important. The ratio is greater that 1 due to the coherence condition, which allows for a larger momentum transfer from the iron nuclei than in gold.  Both ratios drop at very large values of $Q^2$ near the kinematic limits as.  These are reached for gold before iron as the smaller nucleus allows for a higher maximum photon energy.
\section{Discussion and conclusions}
We have calculated the kinematics for photoproduction and electroproduction at an electron-ion collider for a variety of VMs.  The rates are very high for light mesons, and even for the J/$\psi$.  For eRHIC and JLEIC, $\Upsilon$ photoproduction will produce a useful sample, but the rates for electroproduction are moderate, inadequate for detailed studies with multi-dimensional binning.  At LHeC, the rates for all mesons are large. \\
  The kinematic distributions are important in determining the requirements for EIC detectors.  VM photoproduction occurs at a wide range of rapidities.  To study production at low photon energies, requires a detector that is sensitive at large negative rapidities, while studies of production at high photon energies, near the kinematic limits, requires a detector that is sensitive at large positive rapidities.   The need for a large acceptance detector may be partially offset by taking data at multiple collision energies.\\
We thank Joakim Nystrand, Janet Seger, Joey Butterworth and Yury Gorbunov for their work with STARlight, which provided the code base to start this work.   This work was funded by the U.S. Department of Energy under contract number DE-AC-76SF00098.

\bibliography{dis_biblio.bib}

\end{document}